\documentclass[11pt]{article}

\usepackage[utf8]{inputenc}
\usepackage{geometry}
\usepackage{tabularx}
\usepackage{setspace}
\usepackage{amssymb}
\usepackage{hyperref}
\usepackage{booktabs}
\usepackage{xcolor}
\usepackage{graphicx}
\usepackage{pifont}
\usepackage{textcomp}
\usepackage{epstopdf}
\usepackage{subfig}
\usepackage{amsfonts, amsmath, mathtools}
\mathtoolsset{showonlyrefs=true}

\usepackage{authblk}
\usepackage[numbers]{natbib}  % Bibliografia compatibile con arXiv

\title{Modelling benthic animals in space and time using Bayesian Point Process with cross validation: the case of Holoturians}

\author[1]{Daniele Poggio}
\author[2]{Gian Mario Sangiovanni\thanks{Corresponding author: \texttt{gianmario.sangiovanni@uniroma1.it}}}
\author[1]{Gianluca Mastrantonio}
\author[2]{Giovanna Jona Lasinio}
\author[3]{Edoardo Casoli}
\author[4]{Stefano Moro}
\author[3]{Daniele Ventura}

\affil[1]{Department of Mathematical Sciences ‘G.L.Lagrange’, Politecnico di Torino, Corso Duca degli Abruzzi 24, 10129 Torino, Italy}
\affil[2]{Department of Statistical Sciences, University of Rome ‘La Sapienza’, P.le Aldo Moro 5, 00185 Rome, Italy}
\affil[3]{Department of Environmental Biology, University of Rome ‘La Sapienza’, P.le Aldo Moro 5, 00185 Rome, Italy}
\affil[4]{Department of Integrative Marine Ecology (EMI), Stazione Zoologica Anton Dohrn, Via Gregorio Allegri 1, 00198 Rome, Italy}

\date{}

\begin{document}

\maketitle

\begin{abstract}
Understanding the spatial distribution of Holothurians is an essential task for ecosystem monitoring and sustainable management, particularly in the  Mediterranean habitats. However, species distribution modeling is often complicated by the presence-only nature of the data and heterogeneous sampling designs. This study develops a spatio-temporal framework based on Log-Gaussian Cox Processes to analyze Holothurians' positions collected across nine survey campaigns conducted from 2022 to 2024 near Giglio Island, Italy. The surveys combined high-resolution photogrammetry with diver-based visual censuses, leading to varying detection probabilities across habitats, especially within Posidonia \textit{oceanica} meadows. We adopt a model with a shared spatial Gaussian process component to accommodate this complexity, accounting for habitat structure, environmental covariates, and temporal variability. Model estimation is performed using Integrated Nested Laplace Approximation. We evaluate the predictive performances of alternative model specifications through a novel k-fold cross-validation strategy for point processes,  using the Continuous Ranked Probability Score.   Our approach provides a flexible and computationally efficient framework for integrating heterogeneous presence-only data in marine ecology and comparing the predictive ability of alternative models.
\end{abstract}


% --- INIZIO CORPO ARTICOLO ---

% (Incolla qui il contenuto del paper)
\section{Introduction}

Sea cucumbers (Holothuroidea) are ecologically significant marine invertebrates that contribute fundamentally to nutrient cycling, sediment bioturbation across various marine ecosystems, functioning across a range of marine habitats, from shallow seagrass beds to deep‑sea ecosystems\cite{lopez1987ecology, purcell2016ecological, schneider2013inorganic}. Their feeding and bioturbation activities enhance sediment oxygenation and organic matter turnover, rendering them “ecosystem engineers” in many coastal systems. Despite their ecological importance and increasing commercial exploitation \cite{gonzalez2014assessment, hamel2022global}, Mediterranean holothurian populations remain understudied, with significant knowledge gaps regarding their distribution patterns, habitat preferences, and ecological dynamics \cite{pasquini2022new, rakaj2024mediterranean}. This gap is striking given that Mediterranean sea cucumber fisheries have expanded in recent decades, often relying on rudimentary catch records and opportunistic surveys.

Modelling sea cucumber distributions—and marine ecological patterns more broadly—faces two persistent challenges: the heterogeneity of environmental covariates and the presence-only nature of observational data \cite{warton2010poisson}. Environmental variables are typically provided as GIS rasters or spatial lattices, requiring careful preprocessing and alignment before statistical analysis \cite{simpson2016going}. At the same time, records of sedentary benthic invertebrates lack absence data due to their nature, introducing complex detection biases and uneven sampling effort across surveys \cite{yuan2017point, martino2021integration, sicacha2021accounting}. Failing to address these issues can lead to biased inference and suboptimal predictive performance.

In this study, we integrate presence data from nine survey campaigns conducted between $2022$ and $2024$ at Punta Lazzaretto (Giglio Island). The $2023$ campaign, carried out within a Posidonia \textit{oceanica} meadow, employed a distinct sampling technique, further increasing the heterogeneity of the data set. To reconcile these differences, we model the observations of each campaign as a realization of a log‑Gaussian Cox process (LGCP), yielding a space–time Poisson point process\cite{brix2001spatiotemporal, serra2014spatio}. Estimation of parameters has been carried out in a Bayesian framework using the Integrated Nested Laplace Approximation (INLA) via the R-INLA package \cite{rue2009inla}. Drawing on prior studies, we confirmed that a shared Gaussian process effectively captures the underlying spatio-temporal variation.

Our contributions are fourfold. First, this work represents a comprehensive, spatiotemporal LGCP analysis of Mediterranean sea cucumbers that integrates multiple heterogeneous survey protocols. Second, we extend the point‑process cross‑validation strategy of \cite{cronie2024cross} to spatially bounded subsets in a spatio‑temporal setup, enabling localized assessment of predictive residuals. Third, we quantify the influence of Posidonia \textit{oceanica} and other environmental drivers on relative intensity, producing high-resolution maps that can guide habitat conservation and fishery management. Finally, we demonstrate how cross-validation for point processes enables rigorous comparison of alternative models and sampling designs, offering a robust alternative to the Deviance Information Criterion (DIC) by facilitating model comparison in a predictive rather than purely explanatory framework.

The remainder of this paper is structured as follows. In \hyperref[sec:2]{Section~\ref*{sec:2}}, we describe the study area, outline the methodology, and highlight differences among the survey campaigns. We also detail the preprocessing of environmental covariates, the data modelling strategy, and the procedure used for model comparison. \hyperref[sec:3]{Section~\ref*{sec:3}} presents the results of the cross-validation analysis across different models, followed by an in-depth evaluation of the best-performing model. Finally, \hyperref[Sec:4]{Section~\ref*{Sec:4}} is dedicated to the discussion.

\begin{figure}[h]
    \centerline{\includegraphics[angle=-90, scale = 0.53]{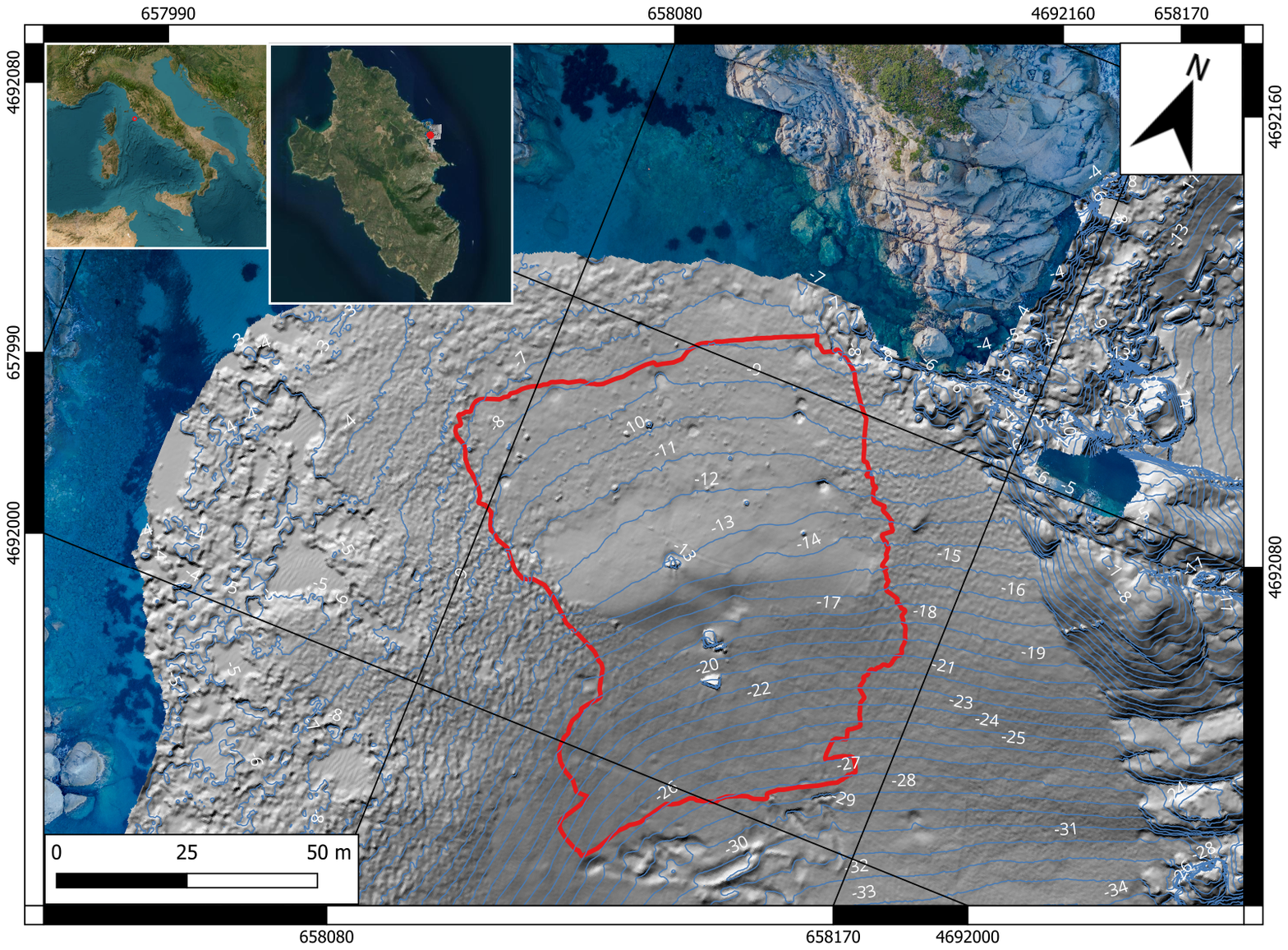}}
    \caption{Map of the study area placed at the north-east coast of the Giglio Island.}
    \label{fig1}
\end{figure}

\section{Materials and Methods}
\label{sec:2}

\subsection{Study Area}
The study site is located at Punta Lazzaretto (N: $ 42.364867$\textdegree; E:$10.920210$\textdegree) on the northeastern coast of Giglio Island in the Tyrrhenian Sea, Italy (Figure \ref{fig1}). It encompasses approximately $5500$ square meters of infralittoral seabed, with depths ranging from $8$ to $27$ meters, characterized by a complex seabed that has undergone significant human-induced disturbances.  This area gained prominence after the Costa Concordia shipwreck incident \cite{casoli2017assessment} since the wreck and associated salvage operations created a shadowing effect that adversely affected the native
\textit{Posidonia oceanica} (P. \textit{oceanica}) Delile meadows, resulting in a marked decline in their coverage \cite{mancini2019impact, toniolo2018seagrass}. After the removal of the wreck in July $2014$ and the completion of seabed remediation in April $2018$, a five-year restoration project began in $2019$, aiming at large-scale
transplantation of P. \textit{oceanica} \cite{mancini2022transplantation}. Currently, the research area exhibits a diverse mosaic of marine infralittoral communities, featuring sandy substrates interspersed with granitic formations, such as boulders and pebbles, which are covered by photophilic algal communities. The site also displays both natural P. \textit{oceanica} patches and a transplanted meadow, colonizing discontinuous patches of dead P. \textit{oceanica} rhizomes with trapped sediments defined as dead "matte" and sandy areas.
% \textcolor{red}{Ma matte ha qualche senso in inglese? o qualche connessione con cosa rappresenta? altrimenti bisogna dargli un altro nome}

\subsection{Data Collection}
\label{Subsection: Data Collection}

\begin{figure}[t]
    \centering
     \subfloat[]{\includegraphics[height=4.8cm, width = 5.8 cm]{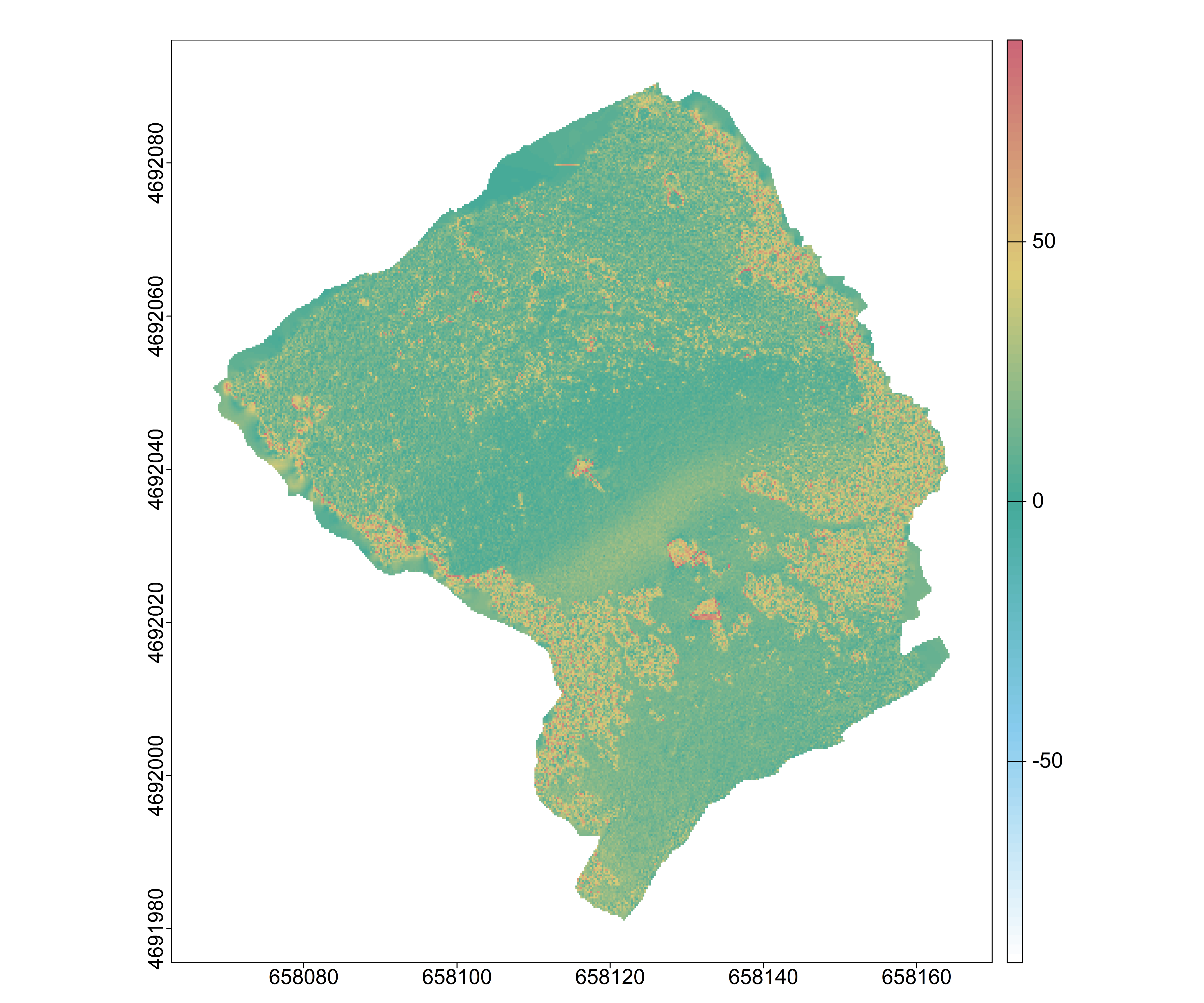}}
     \subfloat[]{\includegraphics[height=4.8cm, width = 5.8 cm]{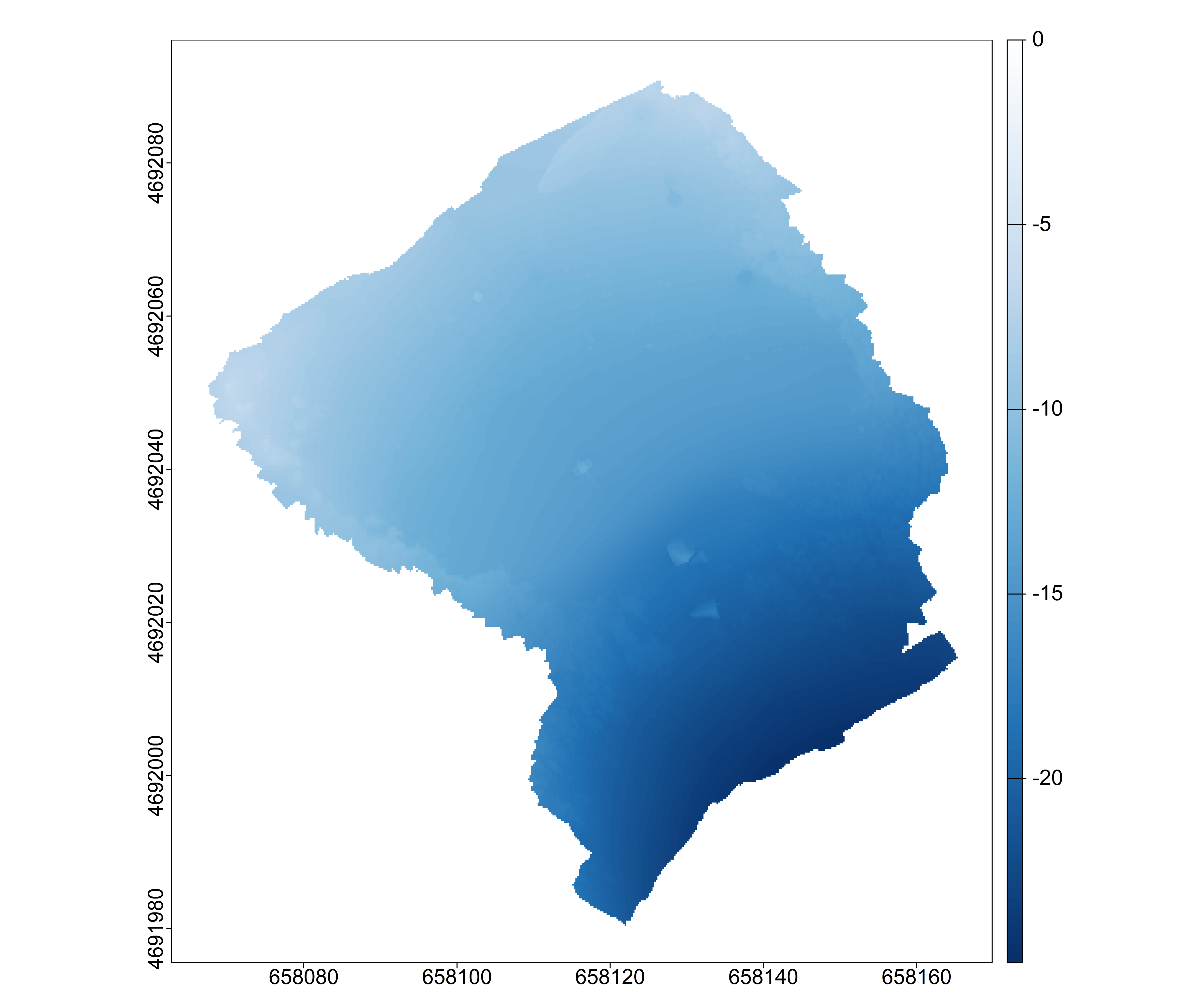}} 
     \subfloat[]{\includegraphics[height=4.8cm, width = 5.8 cm]{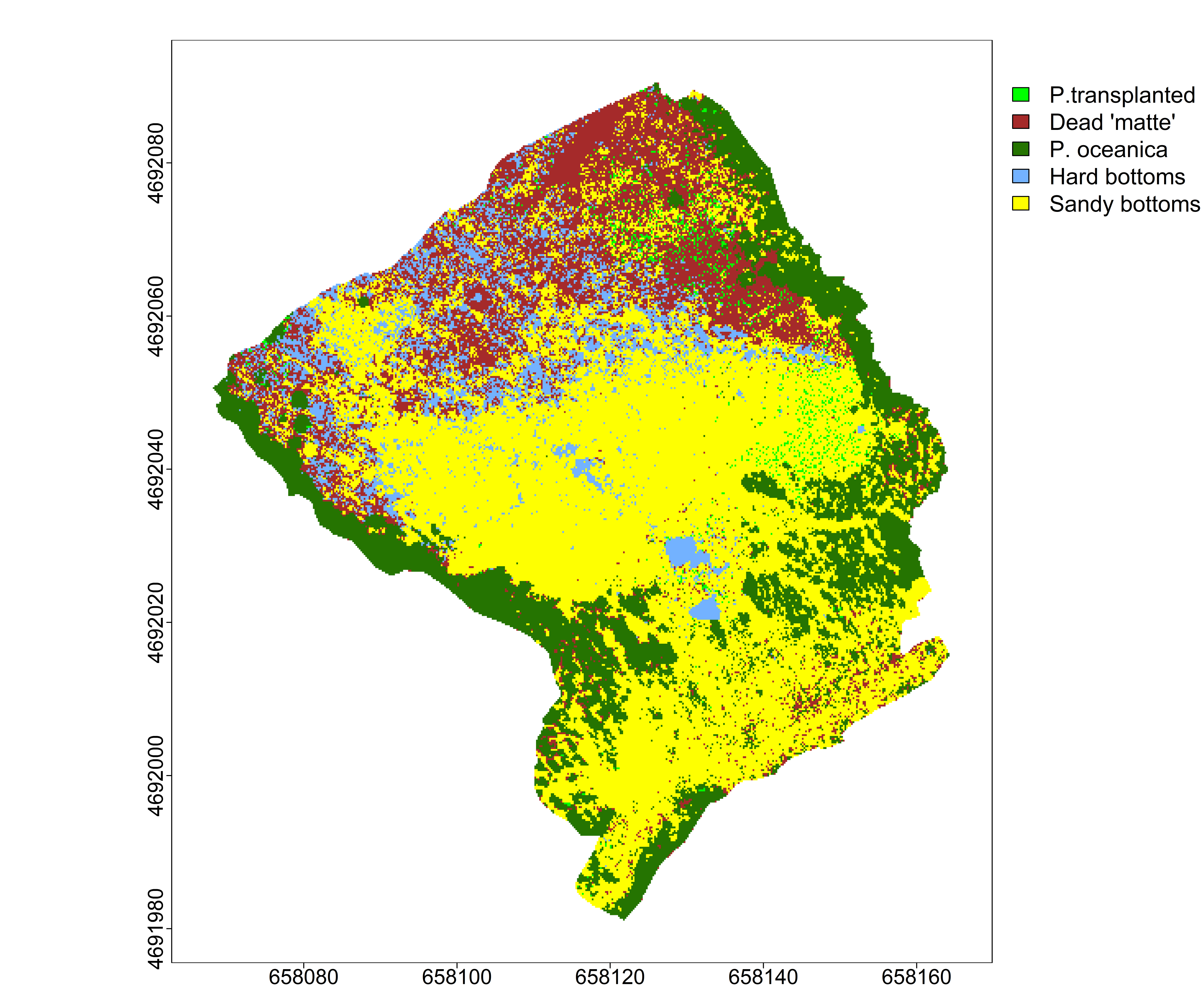}} \\
     \subfloat[]{\includegraphics[height=4.8cm, width = 6.8 cm]{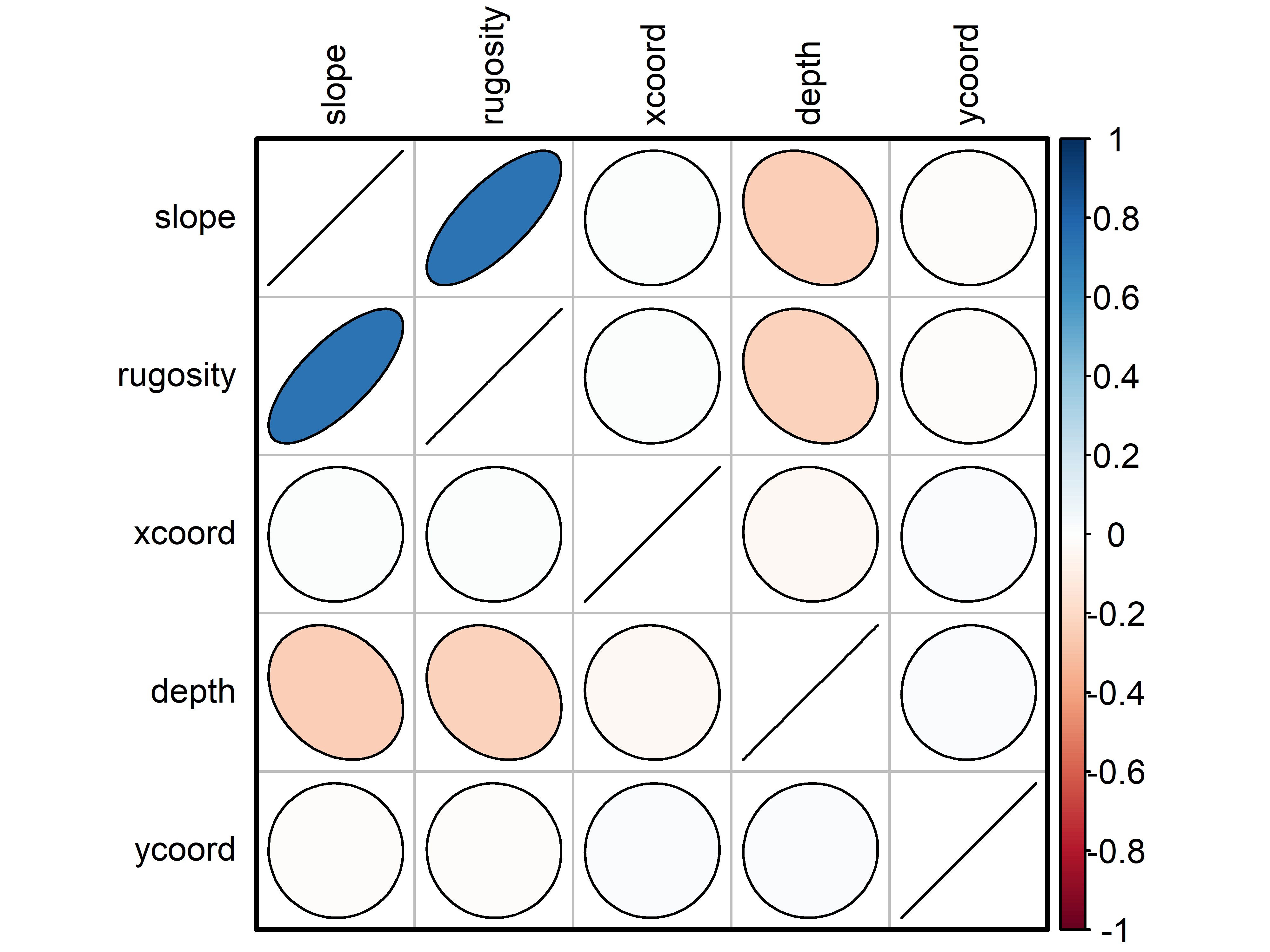}}
     \subfloat[]{\includegraphics[height=4.8cm, width = 5.5
     cm]{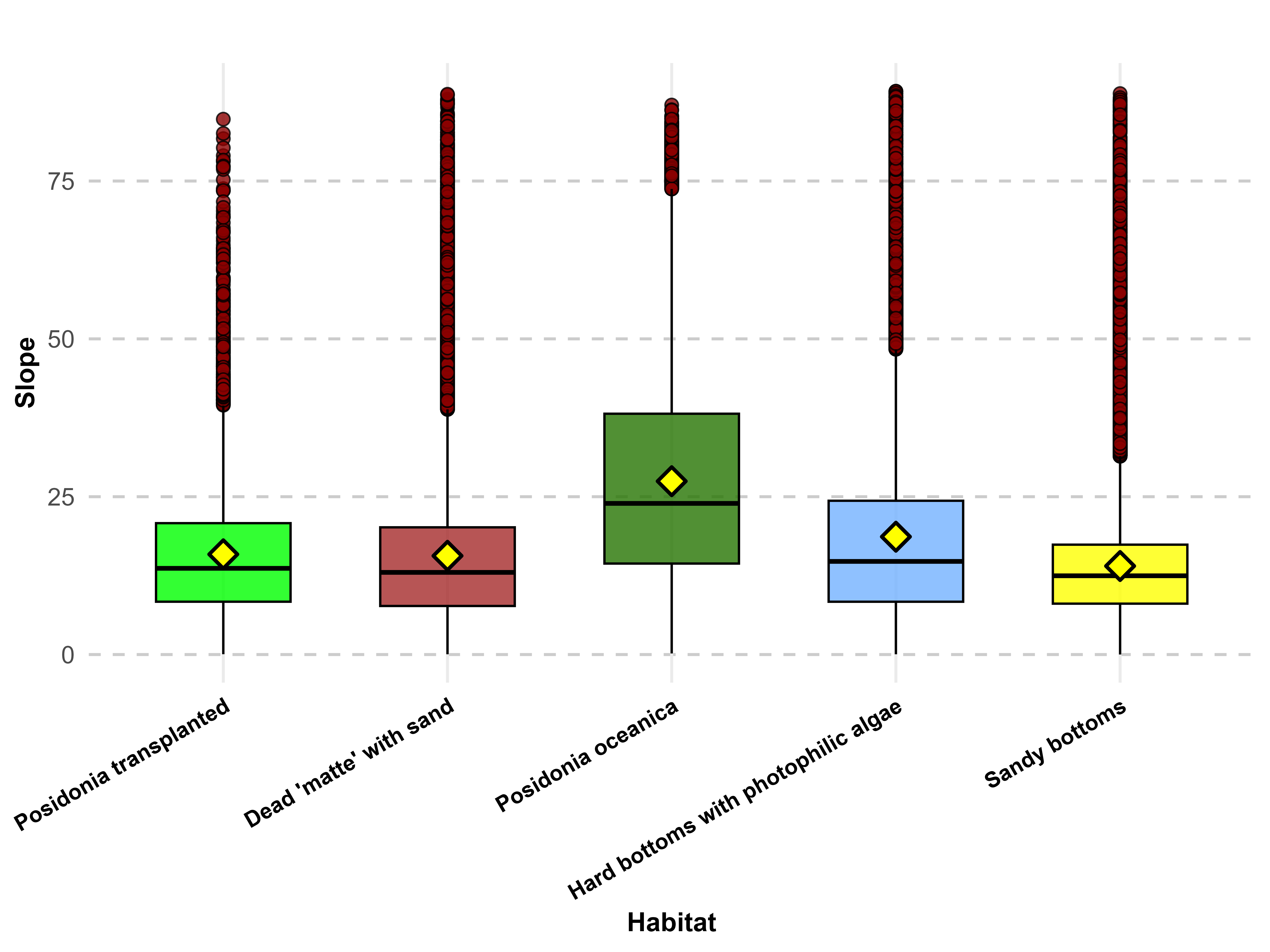}} 
     \subfloat[]
     {\includegraphics[height=4.8cm, width = 5.5 cm ]{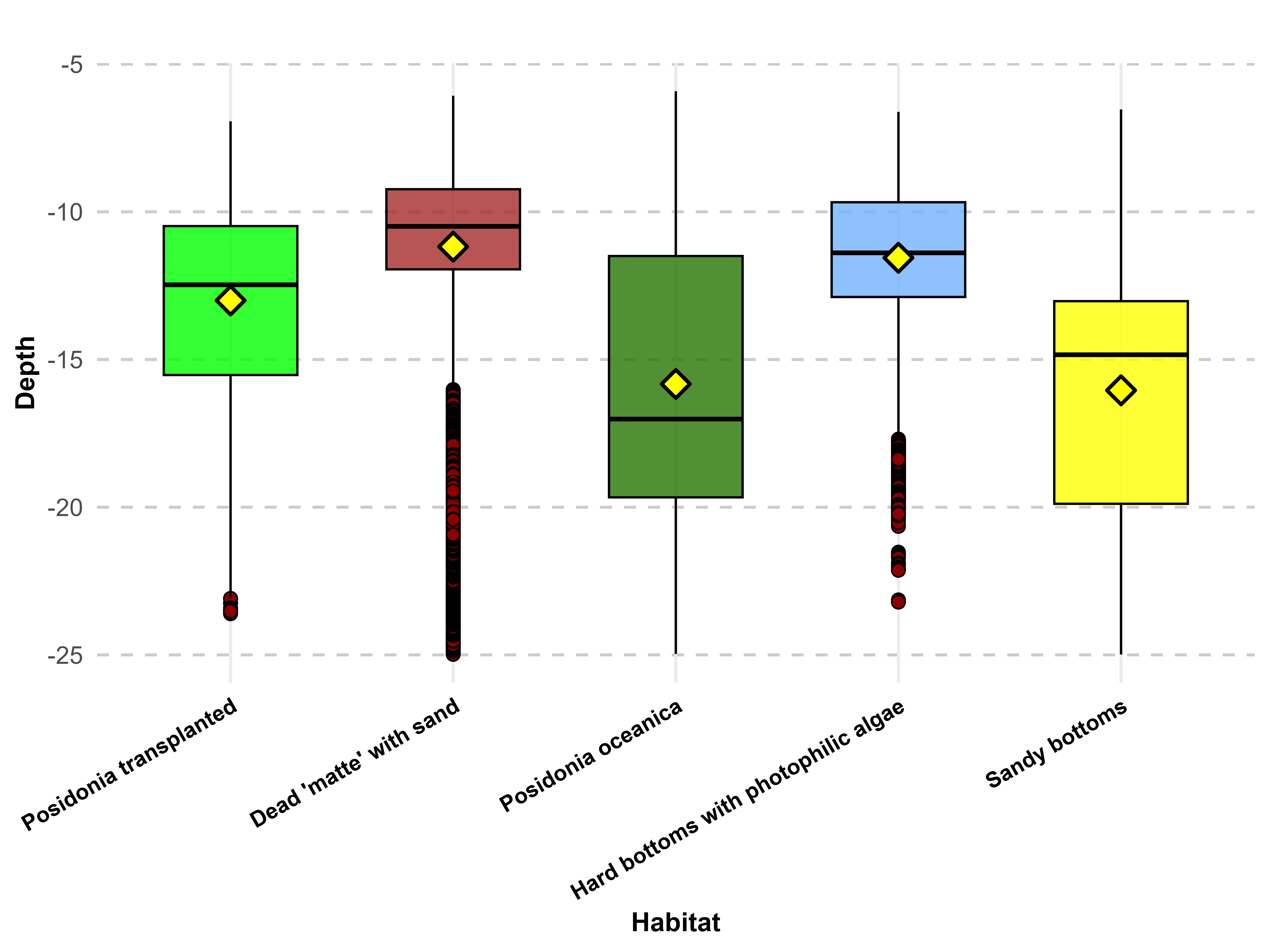}} 
     \caption{Spatial distribution of the variables (a) Slope, (b) Depth and (c) Habitat. The correlation plot of all the covariates is also presented (d). The conditional distribution of slope (e) and depth (d) is presented for the five distinct habitats. The yellow rhomboids represent the value of the covariate mean given a specific habitat .}
     \label{fig: covariates}
   \end{figure}

\begin{figure}[t]
    \centering
     \subfloat[]{\includegraphics[scale = 0.33]{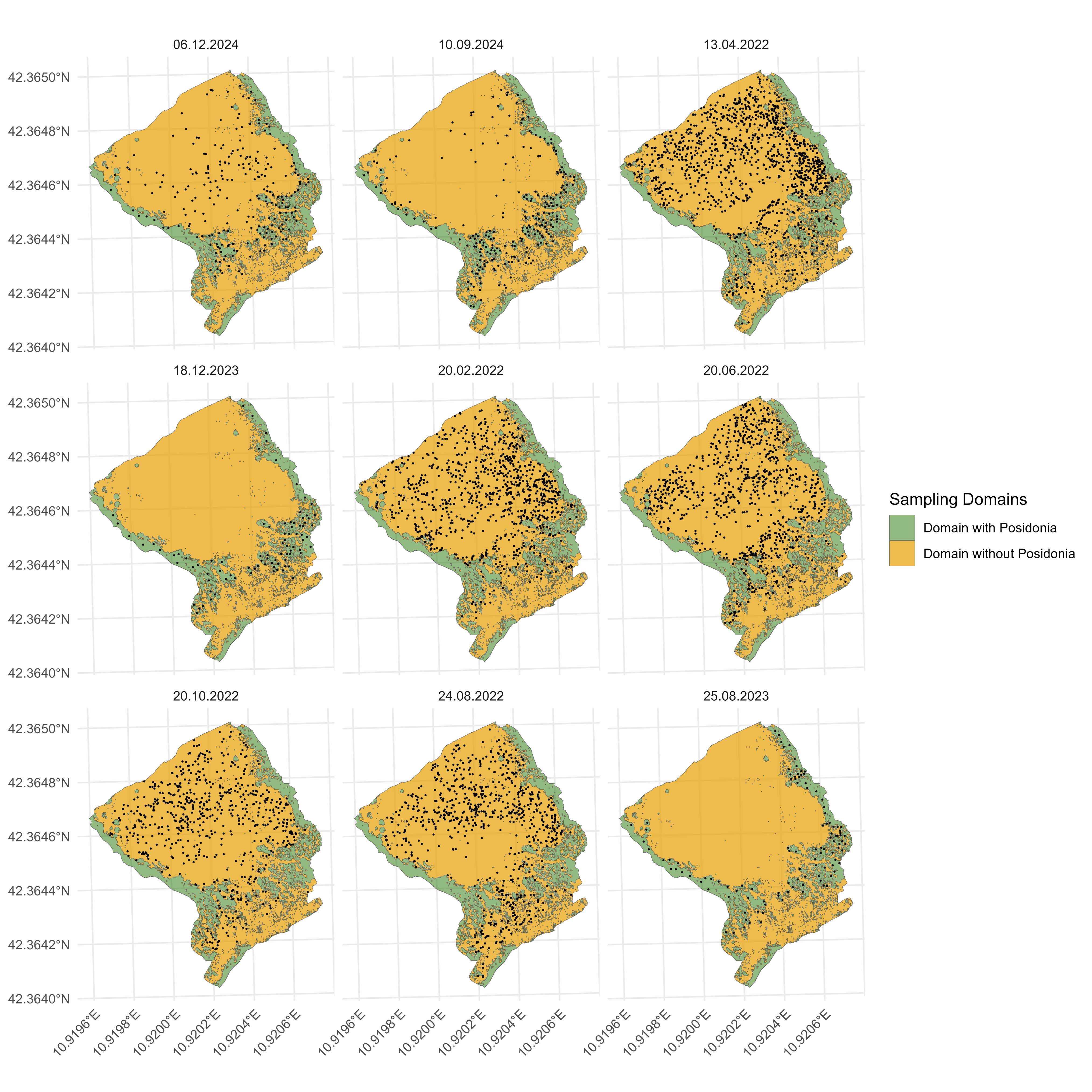}}
     \subfloat[]{\includegraphics[scale = 0.29]{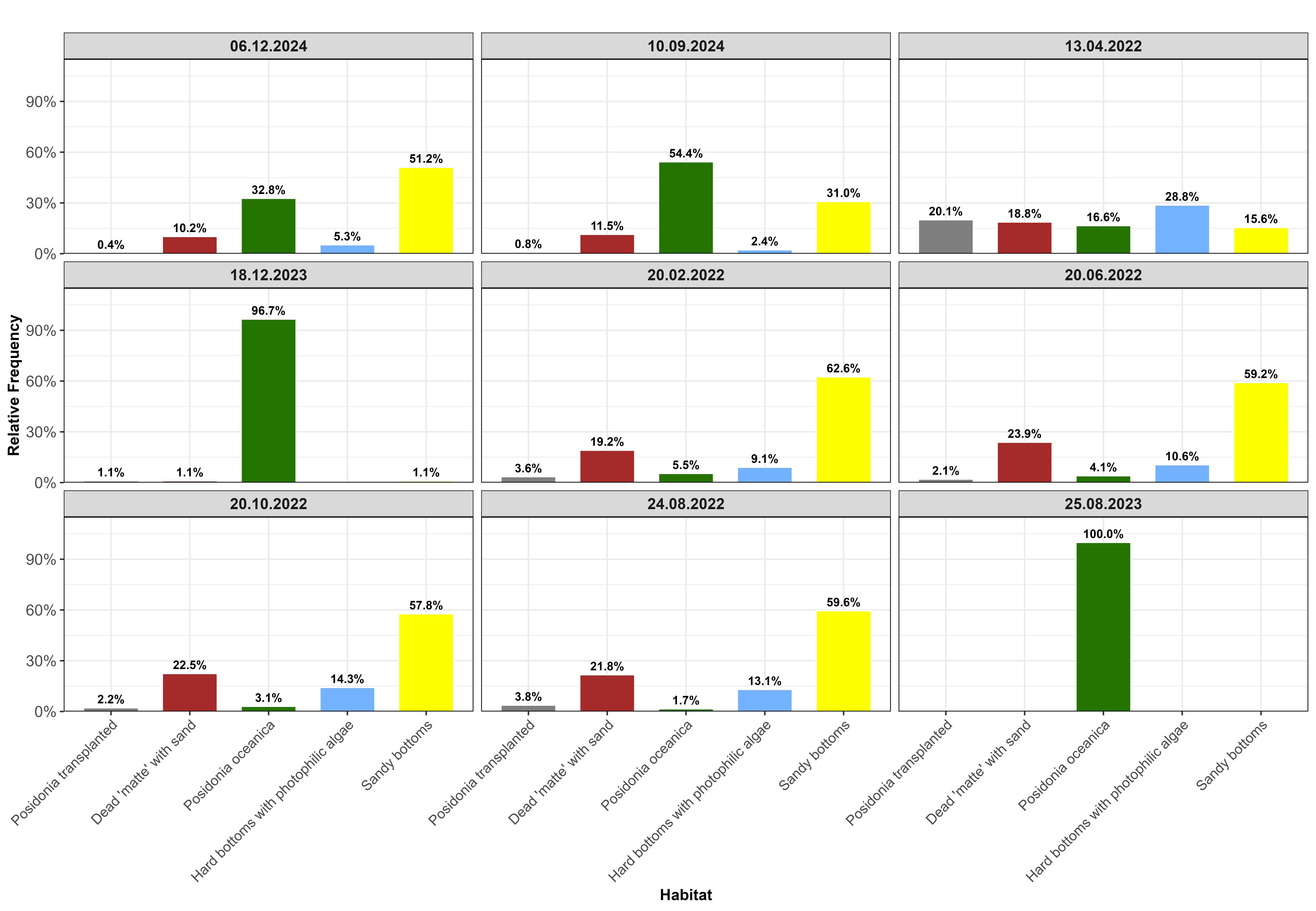}}
     \caption{Spatial distribution of specimens (a) and relative frequencies of observed sea cucumbers within different habitats (b) for the nine sampling campaigns.}
     \label{fig:habitat}
\end{figure}
   
Nine sampling campaigns were carried out in $2022$, $2023$, and $2024$. In $2022$, five campaigns were conducted, and in $2023$ and $2024$, only two campaigns were carried out. 
In $2022$, data were collected using the high-resolution Structure from Motion (SfM) photogrammetric technique in order to monitor the progress of the expansion of restoration initiatives in the study area \cite{ventura2022high} and to assess the structure and composition of benthic communities \cite{ventura2025detecting}. This technique, while effective for broader spatial mapping, may have led to an underestimation of the sea cucumber abundance within natural P. \textit{oceanica} meadows, due to limitations in detecting individuals hidden within dense seagrass. 

To improve the accuracy in estimating the population density within the P. \textit{oceanica} habitat, two additional campaigns were conducted in $2023$. In each campaign of $2023$, three SCUBA divers performed the sampling of sea cucumbers in the natural meadow surrounding the edge of the study area through underwater visual census (UVC). Each diver employed a rope to establish a sequence of linear transects ranging from 50 to 200 meters in length, contingent upon the size of the meadow.
% Images were acquired every 2 s along transects spaced no more than 4 m, ensuring an image footprint of approximately 65 m2 with a ground sample distance of 2–3 mm/pixel
They then conducted a count of specimens within a $1$-meter width, measuring $50$ centimeters from the rope on either side. To identify all sea cucumber specimens, divers manually displaced the seagrass leaves surrounding the rope and documented the estimated locations on a waterproof plasticized map of the study area, which was created using orthophoto mosaics based on SfM obtained previously. When the P. \textit{oceanica} patches exhibited a polygonal shape, particularly in the deeper sections of the study area, the circle search pattern technique was implemented to enhance the manual identification of sea cucumbers. This method involved a diver swimming in concentric circles, progressively enlarging the radius around the center of the patch. A graduated rope was employed to ensure precise control over the distance of each circular trajectory from the center. 

In $2024$, the two campaigns were carried out collecting data using both SfM and UVC, to get estimations of the abundance of sea cucumbers in all the study area, including in the natural P. \textit{oceanica} meadow. The integration of manual and photogrammetric data facilitated the development of a unified analytical framework, combining the strengths of both data collection methods. This comprehensive sampling strategy forms the basis for the results presented in Figure~\ref{fig:habitat} (b). Unsurprisingly, the $2022$ campaigns contributed the highest number of observations, due to the extensive spatial coverage of the SfM approach. However, nearly 90\% of these observations were located outside P. \textit{oceanica} meadows, reflecting a potential habitat bias inherent to the initial sampling design. After a critical quality assessment of the data, observations within P. \textit{oceanica} from $2022$ were excluded due to unreliability, primarily stemming from visibility limitations. Similarly, observations outside of P. \textit{oceanica} from $2023$ were also discarded, as they lacked sufficient observational rigor. This careful filtering process resulted in a clean and well-separated dataset, both spatially and temporally, as summarized in Table~\ref{tab:season_habitat_totals} and Figure~\ref{fig:habitat} (a). This refined dataset not only enhances the robustness of our model inference but also ensures that habitat comparisons are not confounded by sampling methodological inconsistencies across years.

In terms of covariates, we used the same data considered in \cite{ventura2025detecting} and \cite{mastrantonio2024species}, with a resolution of $0.21 \times 0.24$ $m$. We considered continuous spatial covariates such as slope, depth, and rugosity, as well as a categorical covariate representing the benthic habitat type. Depth, slope, and rugosity were derived from high-resolution Digital Surface Models (DSMs).
Due to the extremely high resolution of the imagery, to extract the benthic habitat cover types \cite{fallati2024combining},  an object-based image analysis (OBIA) approach was employed to facilitate the classification of orthophoto mosaics. The classification process involved segmenting the images into homogeneous regions based on spectral and geometric features, followed by a supervised classification that assigned each segment to one of five benthic habitat classes: sandy bottoms, hard bottoms with photophilic algae, dead matte with sand, natural P. \textit{oceanica meadow}, and transplanted P. \textit{oceanica}. Rugosity was excluded from the analysis due to its high correlation with slope. As the quantitative covariates violated the assumption of normality, differences across habitat types were assessed using the Kruskal–Wallis test. For both slope and depth, the test revealed statistically significant differences among habitat classes at the $0.05$ significance level, leading to the rejection of the null hypothesis of homogeneity. A visual comparison of the covariates across habitats is presented in Figure~\ref{fig:habitat}. These environmental variables were aggregated over a regular $1 \times 1 \,\, m$ grid using zonal average due to computational efficiency.

\subsection{Data Modelling} 
We envision the Holoturian positions $\mathbf{U}_t \in \mathcal{D}$, where $t = 1, \dots, 9$ denotes the campaign where the data has been recorded, as a realization of a Log-Gaussian Cox Process (LGCP) \cite{moller1998log}. The spatial domain $\mathcal{D} \subset \mathbb{R}^2$, where the point process is defined, is the union of the P. \textit{oceanica} habitat $\mathcal{D}_{1} \subset \mathbb{R}^2$, while the remaining habitat types are indicated as $\mathcal{D}_{2} \subset \mathbb{R}^2$. The two domains are disjoint, i.e., $\mathcal{D}_{1} \cap \mathcal{D}_{2} = \varnothing$, and their union covers the entire study area: $\mathcal{D}_{1} \cup \mathcal{D}_{2} = \mathcal{D}$.
Let $\lambda_t(\mathbf{s})$ be the intensity function of the LGCP at time $t = 1, \dots, 9$, and spatial location $\mathbf{s} = (s_1, s_2)^T \in  \mathcal{D}$. Each time point $t$ corresponds to a specific sampling campaign indexed as $C_{y. i}$, where $y \in \{1, 2, 3\} $ denotes the year and $i$  the campaign number within that year. Specifically, $ C_{1.1}$  to  $C_{1.5}$ represent the five campaigns conducted in $2022$, $ C_{2,1}$  and $C_{2.2}$ those from $2022$, $C_{3.1}$ and $C_{3.2}$ the two campaigns from $2023$.

The log-intensity $\log(\lambda_t(\mathbf{s}))$ is modelled as a linear combination of parameters $ \mu_{t}$ accounting for changes in the total number of observed individuals across time (campaign effects), a common intercept across seasons $\mu_0$, a fixed term $\beta$ to model the effects of spatial covariates $\mathbf{x}(\mathbf{s})^T = (x_1(\mathbf{s}), \dots, x_p(\mathbf{s}))^T$, an additional term $\gamma$ to take into account the effect of the habitat P. \textit{oceanica} versus the others, and a spatial Gaussian Process (GP) $w(\mathbf{s})$ to account for the spatial correlation in the data. For the campaign effects, we define a second hierarchical level where $\mu_{t}$ is distributed as a Gaussian distribution with $0$ mean and variance $\tau^2$.
The model is defined as follows:
\begin{align}
\mathbf{U}_t &| \lambda_t(\mathbf{s}) \sim \mbox{PP}(\lambda_t(\mathbf{s})) \\
\log(\lambda_{t}(\mathbf{s})) &= \mu_{0} + \mu_{t} + \gamma z(\mathbf{s}) + \mathbf{x}(\mathbf{s})^T \boldsymbol{\beta} + w(\mathbf{s}),\\
w(\mathbf{s})& \sim \mbox{GP} \left(0, C(\cdot; \sigma^2, \rho)\right),\\
\mu_t & \sim N(0, \tau^2).
\label{eq:intensity}
\end{align}
where $\sigma^2$ and $\rho$ are the variance and the range of the Matèrn covariance function, while $\alpha = 2$ \cite{lindgren2011gmrf}.
In this work, $ \mathbf{s} \in \mathcal{D}_{2}$ for $t \in \left\{1,2,3,4,5\right\}$, $\mathbf{s} \in \mathcal{D}_{1}$ if $t \in \left\{6,7\right\}$ and $\mathbf{s} \in \mathcal{D}$ if $t \in \left\{8,9\right\}$.  We adopt this modeling framework for several reasons. First, the term $\mu_t$ allows us to flexibly account for seasonal or campaign-specific differences in the total abundance, without imposing a fixed trend or requiring homogeneous effort across time. This is crucial in ecological studies where survey efforts and environmental conditions vary over campaigns. Second, the inclusion of the binary habitat indicator $z(\mathbf{s})$ and its associated parameter $\gamma$ enables the model to adjust for heterogeneous sampling intensity and ecological characteristics specific to the P.\textit{oceanica} habitat, especially considering its enhanced exploration during certain campaigns. This additive structure for the log intensity leads to a multiplicative decomposition of the intensity function.
\begin{align}
\label{eq:divided_intensity}
    \lambda_{t}(s) = \exp(\mu_{0} + \mathbf{x(s)}^T \mathbf{\beta} + w(\mathbf{s})) 
 \cdot \exp(\mu_{t}) \cdot \exp(\gamma z(s)), \quad s\in \mathcal{D}
\end{align}
In the formulation of Equation \eqref{eq:divided_intensity}, $\exp(\mu_{t})$ accounts for variation across sampling times. In contrast, $\exp(\gamma z(\mathbf{s}))$ corrects the intensity for differing sampling intensities within the P.\textit{oceanica} region ($\mathcal{D}_{1}$), so it can be interpreted as a measure of the sampling effort. This factorization thus provides an effective means of correcting the overall intensity to incorporate these influences.

\subsection{Model Estimation}

For the estimations of some alternative models we used INLA \cite{rue2009inla} \href{https://www.r-inla.org/home}{R-INLA}, and in particular the wrapper \emph{inlabru} \cite{bachl2019inlabru}. INLA is an algorithm for approximate Bayesian inference in Latent Gaussian Models (LGM), where the latent field is Gaussian and governed by a small number of hyperparameters, and the response variables can be non-Gaussian. It relies on the Laplace approximation to efficiently compute posterior distributions.

INLA supports spatial models where the GP is approximated by a Gaussian Markov Random Field using the SPDE approach \cite{lindgren2011gmrf}, which substantially reduces the estimation time. The core idea is to construct a finite-dimensional representation of the spatial Gaussian process over a mesh defined on the spatial domain, and to approximate the LGCP likelihood in a way that makes it tractable for INLA-based inference \cite{simpson2016going}.

{In terms of prior distributions, for the intercept $\mu_0$ and the regression coefficients $\mathbf{\beta}$, we employ non-informative Gaussian priors centered at $0$ and with precision $0.001$. For the parameters $\sigma^2$ and $\rho$ of the spatial Gaussian process, we choose PC priors \cite{simpson2017penalising}. Specifically, we set $\mathbb{P}(\rho < 50) = 0.5$ and $\mathbb{P}(\sigma > 0.5) = 0.01$. Those choices are consistent with the spatial extent of the study area. We set for the inverse of $\tau^2$, i.e. precision, as a Log-Gamma prior with shape and rate fixed to $1.0$ and $0.01$ respectively.}

\subsection{Model Comparison}\label{Model-Comparison}
% scopo: sviluppare qualche metrica per confrontare i diversi modelli e sceglierne
% scartiamo DIC perchè altrimenti non ha senso l'articolo
% decidiamo di utilizzare i raw residual come indicatore della capacità di performance predittiva di ciascun modello, valutando i raw residuals su una griglia spaziale FOR FIXED t
% utilizziamo la cross-validation per calcolare i raw residuals utilizzando gli stessi dati con i quali sono stati fittati i modelli (come vignetta inlabru)
% utilizziamo CRPS e RMSE per valutare distribuzione a posteriori ottenuta sui residui

To effectively compare the performance of alternative models, it's essential to employ evaluation metrics that account for both model complexity and prediction capability. While the DIC is a commonly used tool for model evaluation in Bayesian settings, it has notable limitations \cite{mcelreath2018statistical}. Specifically, the DIC tends to favor overfitted models and lacks consistency in model identification, especially in hierarchical settings. Given these concerns, we opted to asses models within the predictive framework \cite{leininger2017bayesian, gelfand2018bayesian}. In particular, we employ a $K$-fold cross-validation procedure for point processes \cite{cronie2024cross}, allowing us to assess the out-of-sample predictive performance.  As our primary measure of predictive accuracy, we use raw residuals \cite{baddeley2005residual}. To enhance the spatial resolution and adaptability to temporal settings, we extend this approach by partitioning the study region into bounded spatial subsets $B_{1}^{t}, \ldots, B_{G_{t}}^{t}$ at each time point $t$, enabling a localized approximation of the residuals. This partitioning is motivated by the spatial structure of our data: the first five sampling times are defined over $\mathcal{D}_{2}$, the next two over $\mathcal{D}_{1}$, and the final two over the entire region $\mathcal{D}$. As a result, the same partition is used for the $2022$ sampling campaigns, a different one for the $2023$ sampling campaigns, and yet another for the remaining two campaigns. For each time $t$, suppose that $\mathbf{U}_{t}|\lambda_{t}$ is a Non-Homogeneous Poisson Process with intensity function $\lambda_t$ defined over the corresponding spatial domain. A realization of this process is denoted by $\mathbf{u}_t$, which can be partitioned into $\mathbf{u}_t^1, \ldots, \mathbf{u}_t^{G_{t}}$, representing the observed points within each of the bounded subsets. The raw residual in the g-th subset at time $t$ is then defined as
\begin{align}
    \label{eq:residual}
        R_{t}^{g}  = ||\mathbf{u}_{t}^{g}||-  \int_{B_{g}^t}\lambda_t(\mathbf{s}) d\mathbf{s}, 
\end{align}
where $N_{t}^{g} = ||\mathbf{u}_{t}^{g}||$ represents the observed number of points in the g-th bounded subset at time $t$. This residual measures the discrepancy between observed and expected counts, and its expected value is 0 \cite{baddeley2005residual}. The idea behind Cross-Validation for point processes is to construct a marked point process by assigning, at each observed location $\mathbf{u}$, a mark $m(\mathbf{u})$ drawn from a $K$-dimensional Multinomial distribution with equal probability $p = K^{-1}$, where $K$ is the number of employed folds. For each fold $k$, we define the training and testing sets as $\mathbf{u}_{k,t}^{train} = \{u \in \mathbf{u}_t: m(u) \neq k\}$  and $\mathbf{u}_{k,t}^{val} =  \{u  \in \mathbf{u}_t: m(u) =  k\} = \mathbf{u}_t \backslash \mathbf{u}_{k,t}^{train}$, respectively. These are still LGCP with  intensities:
\begin{align}\label{eq:intensity_cv}
    \lambda_t^{train}(\mathbf{s}) =  \frac{K-1}{K}\lambda_t(\mathbf{s}), \quad \lambda_t^{val}(\mathbf{s}) =  \frac{1}{K-1}\lambda_t^{train}(\mathbf{s}).
\end{align}
By using the intensities in (\eqref{eq:intensity_cv}) and partitioning the validation set into $G_{t}$ subsets $\mathbf{u}_{k,t}^{val, g}$ for each time t, we can easily compute the raw residuals for the validation sets
\begin{align}
\label{eq:residual_thin}
    R_{k,t}^{val, g}= || \mathbf{u}_{k,t}^{val, g} || - \int_{{B}^{t}_{g}}\lambda_t^{val}(\mathbf{s}) d\mathbf{s} = || \mathbf{u}_{k,t}^{val, g} || - \frac{1}{K-1}\int_{{B}^{t}_{g}}\lambda_t^{train}(\mathbf{s}) d\mathbf{s} .
\end{align}
An estimator of $R_{k,t}^{val, g}$ is what should be used to assess the model performance, since, if the model is correctly specified, it is expected to have a value close to 0. Because the closed-form expression of the residual posterior distribution is generally unavailable, we use Monte Carlo simulations with a sample size $A$ to approximate it. Evaluating the spatial integral is often the most challenging aspect of this procedure. Since the different domains are in $\mathbb{R}^2$, the integral can be approximated using a quadrature formula based on a discretization of the spatial domain \cite{berman1992approximating}:
\begin{align}
    \label{eq:space_integrale}
    \int_{\mathcal{D}}\lambda^{train}_t(s)ds \approx \sum_{q=1}^Q \alpha_q \lambda_{t}(v_q),
\end{align}
where $\mathcal{V} = \left\{v_q \in \mathcal{D}\right\}_{q = 1}^{Q}$ are the quadrature nodes and $\mathcal{W} = \left\{\alpha_q\right\}_{q = 1}^{Q}$ are the corresponding weights. At the end of this process, we obtain a $A \times K \times G_{t}$ tensor, where $A$ is the number of Monte Carlo replicates and each column contains the residual samples corresponding to each $k$-fold given a subset. If we repeat this procedure for all the $T$ sampling times, we obtain $T$ tensors.  Furthermore, as our objective is to assess the predictive performance of $L$ competing models, denoted by $\mathcal{M}_{1}, \ldots, \mathcal{M}_{L}$, we obtain $L \times T$ different tensors. To have a fully Bayesian evaluation of the model performance, we can use proper scoring rules, which provide summary measures to evaluate probabilistic forecasts by assigning a numerical score \cite{gneiting2007strictly}. In particular, we opted for the Continuous Ranked Probability Score (CRPS)\cite{matheson1976scoring}, measuring the mean square error between the predicted and empirical cumulative distribution functions. In a general setting where we want to predict the value of an observation $y$, it can be expressed as 
\begin{align}
    \label{eq:CRPS}
    \text{CRPS}(F, y) = \int_{\mathbb{R}} (F(x) -\mathbb{1}{\{ x \geq y\}})^2 dx
\end{align}
where $F(\cdot)$ is the predictive distribution of $y$  and  $\mathbb{1}$ is the Heaviside step-function.   In our application, given that we assumed $0$ value for the residual mean, we are particularly interested in the case when $y = 0$. Furthermore, indicating with $F_{R_{k,t}}^{val, g}$ the predictive distribution of the raw residuals, the CRPS is then
\begin{align}
    \label{eq:CRPS2}
    \text{CRPS}(F_{R_{k,t}}^{val, g}, 0) = \int_{\mathbb{R}}  (F_{R_{k,t}}^{val}(x) -\mathbb{1}{\{ x \geq 0\}})^2 dx
\end{align}
The integral in Equation \eqref{eq:CRPS2} cannot be computed analytically, but it can be approximated \cite{kruger2021predictive, jordan2019evaluating} as 
\begin{align}
\label{eq:crps_emp}
\text{CRPS}(F_{R_{k,t}}^{val, g}, 0) \approx \frac{1}{A}\displaystyle\sum_{a = 1}^{A}|r_{k,t}^{g, a}-0| - \frac{1}{2A^{2}}\displaystyle\sum_{a = 1}^{A}\sum_{l = 1}^{A}|r_{k,t}^{g, a} - r_{k,t}^{g, l}|
\end{align}
where $r_{k,t}^{g, a}$ are samples from the distribution $F_{R_{k,t}}^{val, g}$, which can be easily obtained from the posterior marginal distributions of the fitted model by the \emph{inlabru} wrapper. To apply Equation \eqref{eq:space_integrale} for each posterior sample, it is necessary to draw the value of $\lambda_{t}$ from its posterior distribution for all the quadrature points. From a computational standpoint, the computation of residual samples across time points $t$ can be parallelized without loss of generality, significantly improving the efficiency. Lastly, since we employ cross-validation, we compute a more robust estimate of the CRPS by averaging over all folds:
\begin{align}\label{eq:crps_cvw}
\text{CRPS}_{CV}^{g, t} =  \frac{1}{K}  \sum_{k=1}^K \text{CRPS}(F_{R_{k,t}}^{val, g}, 0)
\end{align}
resulting in $G \times T$ CRPS values. In the context of such scores, constructing a local map of residuals/CRPS values becomes a viable solution. This map can offer valuable insights into the model's capacity. Nonetheless, it is possible to aggregate these scores over time and partitions to obtain a single summary score. This then allows for comparisons to be made across the $L$ candidate models. In conclusion, the model with the lower CRPS score is the most effective of the ones examined.

\section{Results}
\label{sec:3}

\begin{figure}[t]
    \centering
     \subfloat[]{\includegraphics[scale = 0.49]{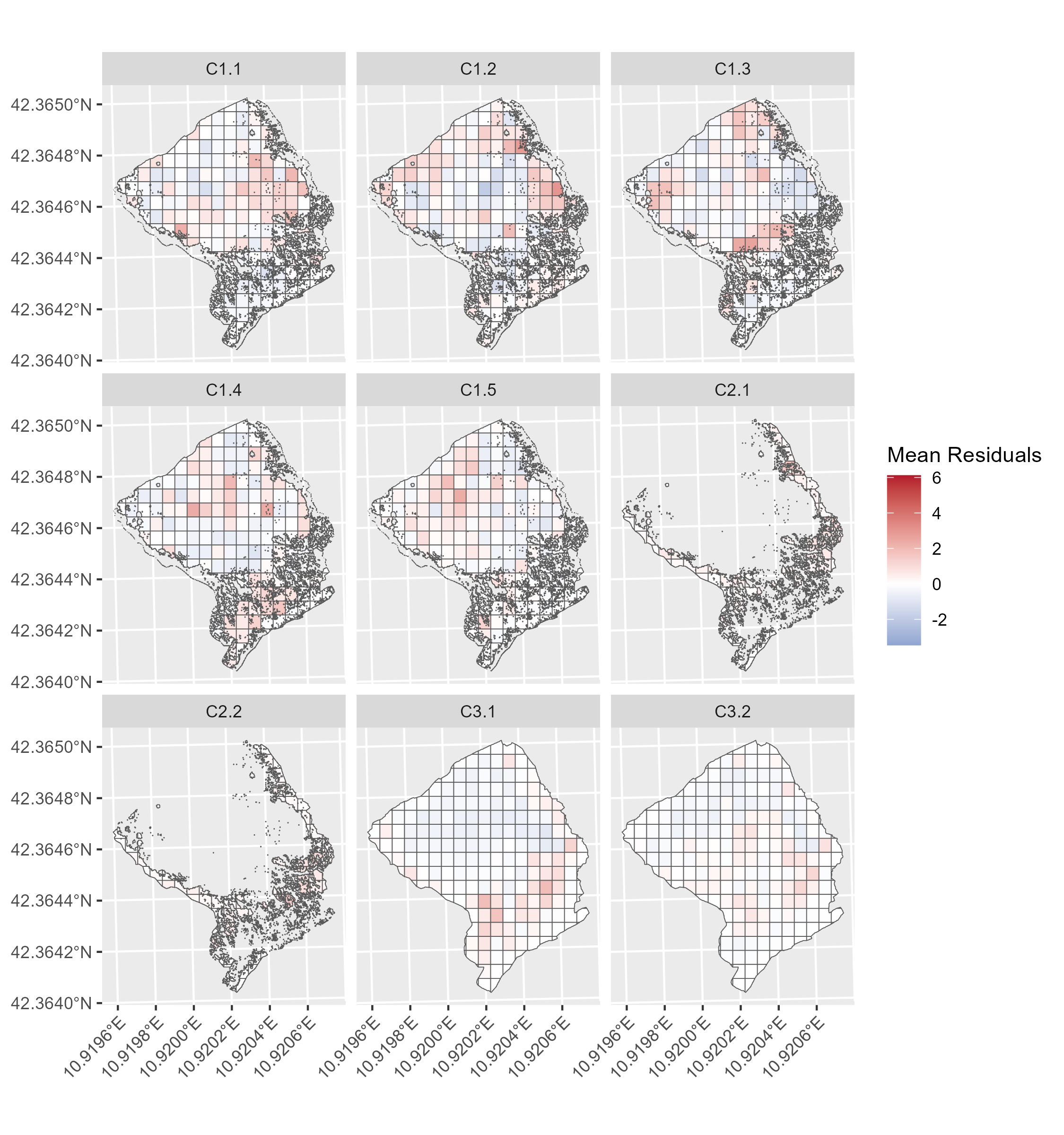}}
     \subfloat[]{\includegraphics[scale = 0.46]{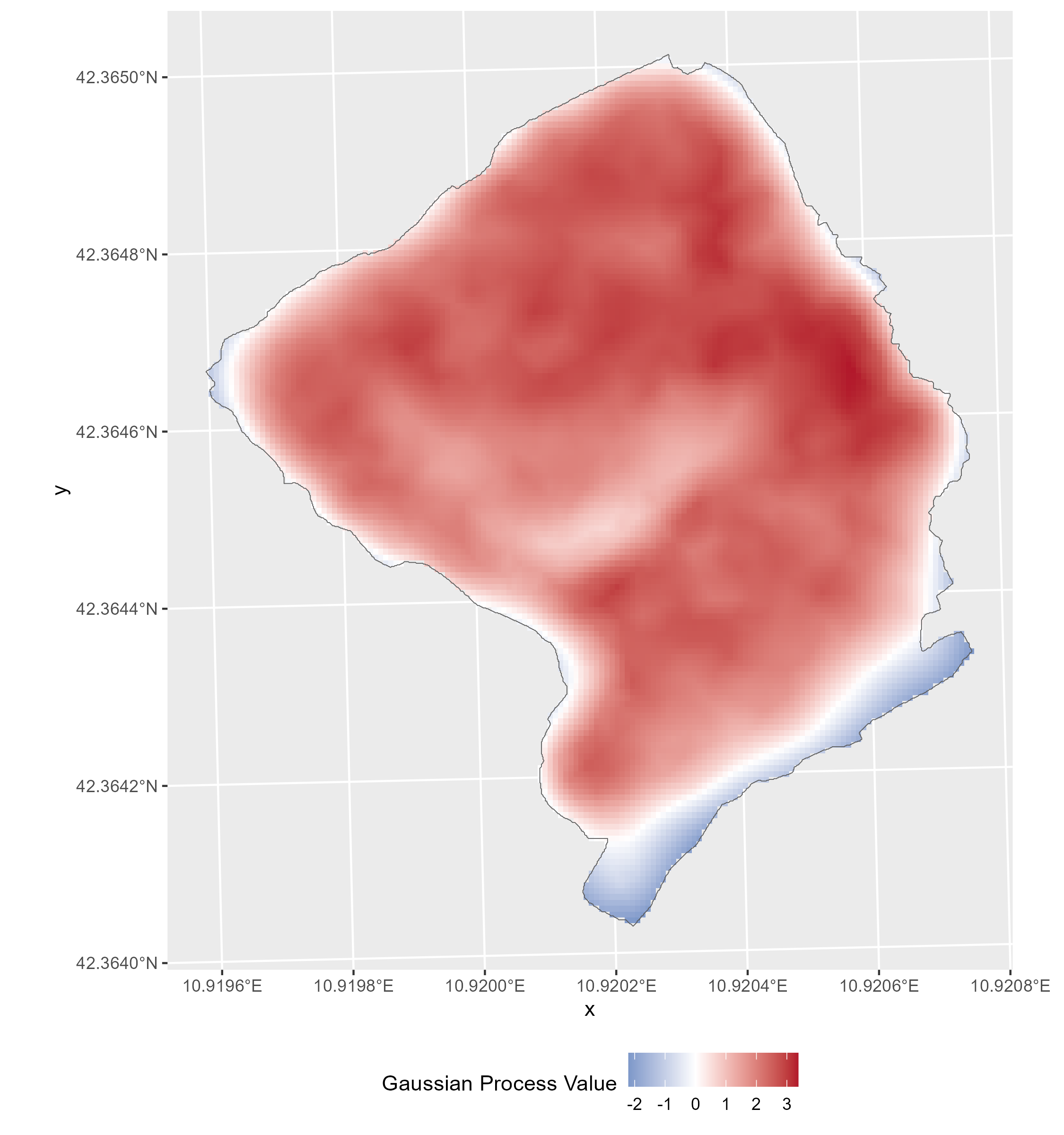}} 
     \caption{Average value of the residuals for the chosen model across the different sampling times and the different defined partitions (a) and Posterior mean for the shared GP (b).}
     \label{fig:residual map}
   \end{figure}

In this section, we present the results of the proposed model applied to the motivating dataset. We tested $240$ models, each of them with a different combination of the following covariates: habitat types, seabed slope, depth, and spatial coordinates. These covariates were selected based on prior ecological knowledge and after exploratory data analysis (Subsection \ref{Subsection: Data Collection}). 
All model combinations were tested under the constraint that P. \textit{oceanica} was always included as a covariate. Due to identifiability constraints, not all habitat types could be included simultaneously as separate indicators.
Given the relatively low number of observations in the most recent sampling campaigns, we opted for a $5$-fold cross-validation scheme to ensure that each fold retained sufficient data for reliable estimation. The group assignments for the folds were generated once and kept fixed throughout all model evaluations to ensure consistency and comparability of the results. The choice of the number of folds in cross-validation can be adapted based on the available data, fewer folds may be preferable in cases of limited observations to avoid overly small training and test sets. However, it is important to consider the computational burden: increasing the number of folds significantly increases the number of model fittings, especially when repeated across multiple replicates and models, as in our study. We set the number of Monte Carlo replicates used for integral approximation to $1000$, and divided each sampling domain into $324$ bounded subsets using an $18 \times 18$ regular grid. This resolution provided a sufficiently detailed representation of the spatial domain while remaining computationally tractable. The grid resolution can be adjusted based on the spatial extent and resolution of the data. %\footnote{The code used for the grid definition, cross-validation implementation, and full results is available in the GitHub repository linked to this article.}.

For each model, we computed residuals and the CRPS. Table \ref{tab: CRPS Score} reports the top $10$ models ranked by CRPS. The first observation is that the CRPS values differ only in the fourth or fifth decimal place, suggesting that all top models perform similarly in terms of predictive accuracy. Given this similarity, we prioritized models where all covariates had $95\%$ credible intervals that did not include zero and whose effects were ecologically interpretable. When using DIC for model selection instead, the model rankings changed substantially. This discrepancy raises questions about the reliability of DIC for model comparison in this context.

Based on these criteria, we selected the model that includes Dead Matte and P. \textit{Transplanted} as habitat covariates, alongside P. \textit{oceanica}. Notably, Sandy Bottoms appears in about half of the best-performing models, but its $95\%$ credible interval consistently includes zero. Neither depth nor slope were retained in the list of the best models. Figure~\ref{fig:residual map} (a) shows the residual mean for the selected model across the nine different sampling campaigns. There does not appear to be a clear spatial pattern in the residuals, in fact the model does not systematically overestimate or underestimate in specific areas.

Table~\ref{tab:posterior_summaries} reports the posterior summary statistics for the intercept, regression coefficients, campaign-specific effects, their precision, and the hyperparameters of the spatial Gaussian process. The intercept is significantly negative, indicating a low baseline intensity. The habitat covariates show distinct effects: P. \textit{oceanica} has a consistently negative coefficient, suggesting reduced expected counts in this habitat relative to the others, possibly due to differential detectability of Sea Cucumbers in this particular habitat. Dead Matte also has a small negative effect, while P. \textit{transplanted} exhibits a positive and well-identified influence on the intensity. The campaign-specific effects reflect considerable variability over time. Campaigns in $2021$  generally show positive values, suggesting higher total abundances relative to the overall mean. In contrast, campaigns from $2022$  have strongly negative effects, highlighting a notable drop in the observed intensity. Campaigns from $2023$ fall in between, with moderately negative effects. These random effects underscore the importance of adjusting for campaign-specific variation in total counts, consistent with the ecological understanding of fluctuations in detection or abundance across seasons and years. The posterior mean of the GP, shown in Figure~\ref{fig:residual map} (b), reveals a clear spatial structure. The GP captures the spatial variation in the intensity of occurrences that is not explained by the covariates or by the campaign-specific effects. The spatial field shows higher values concentrated in the northern and northeastern parts of the study area, indicating regions where the observed point intensity is systematically higher than what would be expected based solely on the included covariates. This could be due to local habitat features not being explicitly modeled. Conversely, the southern and southeastern edge of the domain exhibits lower GP values. These areas are characterized by limited ecological visibility during field campaigns. Furthermore, the posterior estimate of the range parameter implies that on average, spatial correlation decays significantly beyond approximately $56$ meters.  It suggests that the residual variation is locally structured but not globally persistent across the entire domain.

\section{Discussion}
\label{Sec:4}

%\begin{figure}[t]
%    \centering
%    \subfloat[]{\includegraphics[scale = 0.4]{Pictures/lambda_intensity.png}}
%    \subfloat[]{\includegraphics[scale = 0.4]{Pictures/lambda_intensity_q0025.png}}\\
%    \subfloat[]{\includegraphics[scale = 0.4]{Pictures/lambda_intensity_q0975.png}}
%    \caption{For each sampling campaign, the posterior mean (a), the $0.025$ quantile (b), and the $0.975$ quantile (c) of the intensity function are shown.}
%    \label{fig:posterior_lambda}
%\end{figure}

Sustainable management of sea cucumber populations, particularly in the Mediterranean, requires detailed knowledge of species-habitat associations and the consequences of their depletion for ecosystem functioning.  Holothuria tubulosa and related species serve key roles in benthic ecosystems as deposit feeders and bioturbators, influencing sedimentary processes and nutrient cycling \cite{purcell2016ecological}. Despite this, management frameworks often lack robust data, and exploitation continues in many areas, frequently under unregulated or poorly documented conditions \cite{rakaj2024mediterranean}. Our study addresses this gap by combining high-resolution spatial modeling with multiple sampling protocols, enabling fine-scale inference on distributional patterns. In contrast to traditional census methods, which are logistically demanding, our approach integrates photogrammetric surveys with diver-based transect data, accounting for the variability in detection related to habitat structure. This integration is crucial in environments dominated by dense seagrass, where detection is known to be impaired due to visual occlusion and topographic complexity. The model results highlight distinct habitat effects: P. \textit{oceanica} exhibits a consistently negative association with sea cucumber counts, likely due to both lower detectability and potentially less favorable habitat conditions. Dead Matte also shows a negative but weaker effect, suggesting it may offer suboptimal habitat or support lower densities. In contrast, P. \textit{transplanted} areas display a clear positive association with sea cucumber intensity, indicating that these restored habitats may support favorable environmental conditions or increased visibility, thus enhancing detectability and/or actual presence.

In this study, we implemented the k-fold cross-validation procedure for point processes as theoretically developed by \cite{cronie2024cross}, and extended it to spatio-temporal point process models. To the best of our knowledge, this represents the first practical implementation of this method in a spatio-temporal setting. This is a significant contribution that allows for model comparison in a predictive framework, providing a robust alternative to the traditional DIC. We evaluated a total of 240 models, encompassing various combinations of continuous and categorical spatial covariates. The cross-validation framework enabled a systematic comparison based on the  CRPS, but every suitable proper score can be used. Notably, in our case study, this procedure yielded a markedly different ranking of models compared to the DIC, highlighting that DIC may not always be the most appropriate choice for model selection in certain contexts. This observation aligns with findings by \cite{leininger2017bayesian}, who emphasized that model comparisons should be conducted in a predictive space, as traditional criteria like DIC may not adequately capture a model's predictive performance.

Most of the analyses were conducted using standard computational resources, thanks to the efficiency of the INLA algorithm, as implemented in the R-INLA and inlabru packages. To further accelerate computations during the cross-validation phase, we leveraged the Terastat 2.0 infrastructure \cite{9251143}. Specifically, we employed 4 cores with 2 GB of RAM each, achieving an average runtime of approximately 15 minutes per model, which is highly efficient given the complexity of the underlying operations. It is important to note that the cross-validation procedure is method-agnostic and can, in principle, be applied within any inferential framework.

Despite the methodological strengths of our framework, the application to the sea cucumber data revealed only marginal differences in the predictive performance across the tested models. Two factors may explain this outcome. First, the study area may lack sufficient variation in key environmental covariates, limiting the model’s ability to distinguish meaningful patterns in species abundance. Second, the supervised classification process used to generate habitat labels from imagery may have introduced a systematic bias, potentially obscuring ecological signals and reducing the discriminative power of the models.
In summary, our work demonstrates how integrating spatio-temporal modeling with multiple survey modalities can improve ecological inference for benthic species. By explicitly accounting for habitat structure and survey heterogeneity, this approach contributes both methodologically and practically to ecosystem-based fishery assessments and habitat restoration monitoring.

\subsection*{Author contributions}

Poggio Daniele and Sangiovanni Gian Mario contributed equally as first authors.

% This is an author contribution text. This is an author contribution text. This is an author contribution text. This is an author contribution text. This is an author contribution text. 

 \subsection*{Financial disclosure}

Ventura Daniele discloses financial support provided by an entity associated with himself. He also acknowledges a funding relationship involving grants from this same source. All other authors declare no known financial interests or personal relationships that could be perceived to have influenced the content of this work.
% None reported.

\subsection*{Conflict of interest}

The authors declare no potential conflict of interest.

\subsection*{Data availability}

Data are available upon request.

\section*{Acknowledgments}
This research was partially supported by the Pure Ocean Fund (‘3DR-4-Seac’ research grant) and PRIN 2022 (Project: LAGO-ON, grant number 20224LYJM5), which contributed to the acquisition of some equipment used for underwater field sampling. Additionally, Daniele Poggio's participation was made possible through the PNRR-NGEU project, funded by the Ministry of University and Research (MUR) under DM 630/2024. The authors would like to thank Prof. Alessio Pollice for his valuable discussions and insightful suggestions.

\newpage

\begin{table}[t]
    \centering
    \renewcommand{\arraystretch}{1.1}
    \begin{tabularx}{\textwidth}{l|*{5}{>{\centering\arraybackslash}X}|>{\centering\arraybackslash}X}
        \toprule
        \textbf{Sampling Date} & \textbf{P. Transplanted} & \textbf{Dead 'Matte'} & \textbf{\textit{P. oceanica}} & \textbf{Hard Bottom} & \textbf{Sandy Bottom} & \textbf{Total} \\
        \midrule
        20 Feb 2022 & 35 & 189 & 0 & 90 & 616 & 930 \\
        13 Apr 2022 & 302 & 282 & 0 & 433 & 235 & 1252 \\
        20 Jun 2022 & 20 & 227 & 0 & 101 & 562 & 910 \\
        24 Aug 2022 & 30 & 171 & 0 & 103 & 468 & 772 \\
        20 Oct 2022 & 15 & 151 & 0 & 96 & 388 & 650 \\
        25 Aug 2023 & 0 & 0 & 110 & 0 & 0 & 110 \\
        18 Dec 2023 & 0 & 0 & 88 & 0 & 0 & 88 \\
        10 Sep 2024 & 2 & 29 & 137 & 6 & 78 & 252 \\
        06 Dec 2024 & 1 & 25 & 80 & 13 & 125 & 244 \\
        \midrule
        \textbf{Total} & \textbf{405} & \textbf{1074} & \textbf{415} & \textbf{842} & \textbf{2472} & \textbf{5208} \\
        \bottomrule
    \end{tabularx}
    \caption{Sea cucumber observations stratified by season and habitat, excluding records within \textit{P. oceanica} for 2022 surveys and excluding records outside \textit{P. oceanica} for 2023 surveys.}
    \label{tab:season_habitat_totals}
\end{table}

\begin{table}[t]
    \centering
    \renewcommand{\arraystretch}{1.1}
    \begin{tabularx}{\textwidth}{>{\centering\arraybackslash}X
                                  >{\centering\arraybackslash}X
                                  >{\centering\arraybackslash}X
                                  >{\centering\arraybackslash}X
                                  >{\centering\arraybackslash}X
                                  >{\centering\arraybackslash}X
                                  >{\centering\arraybackslash}X
                                  >{\centering\arraybackslash}X
                                  >{\centering\arraybackslash}X}
        \toprule
        \textbf{Depth} & \textbf{Slope}& \textbf{Xcoord} & \textbf{Ycoord} &  \textbf{P. Transplanted} & \textbf{Dead 'Matte'}  & \textbf{Hard Bottom} & \textbf{Sandy Bottom} & \textbf{CRPS} \\
        \midrule
         \ding{54}& \ding{54} & \ding{54} & \ding{54} & \ding{52} & \ding{52} &  \ding{54} & \ding{52} & $0.454320$ \\
         \ding{54}& \ding{54} & \ding{54} & \ding{54} & \ding{52} & \ding{52} &  \ding{54} & \ding{54} & $0.454329$ \\
         \ding{54}& \ding{54} & \ding{52} & \ding{54} & \ding{52} & \ding{52} &  \ding{54} & \ding{54} & $0.454373$\\
         \ding{54}& \ding{54} & \ding{52} & \ding{52} & \ding{54} & \ding{52} &  \ding{54} & \ding{52} & $0.454388$\\
         \ding{54}& \ding{54} & \ding{52} & \ding{52} & \ding{52} & \ding{54} &  \ding{54} & \ding{54} & $0.454389$\\
         \ding{54}& \ding{54} & \ding{54} & \ding{52} & \ding{52} & \ding{52} &  \ding{54} & \ding{52} & $0.454400$\\
         \ding{54}& \ding{54} & \ding{54} & \ding{54} & \ding{54} & \ding{52} &  \ding{54} & \ding{52} & $0.454411$\\
         \ding{54}& \ding{54} & \ding{54} & \ding{52} & \ding{52} & \ding{52} &  \ding{54} & \ding{54} & $0.454416$\\
         \ding{54}& \ding{54} & \ding{52} & \ding{52} & \ding{52} & \ding{52} &  \ding{54} & \ding{52} & $0.454420$ \\
         \ding{54}& \ding{54} & \ding{52} & \ding{52} & \ding{52} & \ding{52} &  \ding{52} & \ding{54} & $0.454451$ \\
        \bottomrule
    \end{tabularx}
    \caption{Covariates included in the top $10$ models ranked by CRPS. Ticks denote inclusion in model fitting computation, while crosses indicate exclusion. The inclusion of a covariate does not imply statistical significance of its effect.}
    \label{tab:CRPS-Score}
\end{table}

\begin{table}[t]
    \centering
    \renewcommand{\arraystretch}{1.1}
    \begin{tabular}{l|rrrrr}
        \toprule
        \textbf{Parameter} & \textbf{Mean} & \textbf{SD} & \textbf{2.5\%} & \textbf{Median} & \textbf{97.5\%} \\
        \midrule
        Intercept & -4.718 & 0.075 & -4.864 & -4.718 & -4.572 \\
        P. \textit{oceanica}    & -0.388 & 0.071 & -0.528 & -0.388 & -0.248 \\
        Dead Matte        & -0.134 & 0.054 & -0.239 & -0.134 & -0.028 \\
        P. \textit{transplanted} &  0.308 & 0.100 &  0.113 &  0.308 &  0.504 \\
        \midrule
        \textbf{Random Effect} & & & & & \\
        \midrule
        C1.1 &  0.819 & 0.035 &  0.749 &  0.819 &  0.888 \\
        C1.2 &  1.116 & 0.032 &  1.053 &  1.116 &  1.179 \\
        C1.3 &  0.797 & 0.036 &  0.727 &  0.797 &  0.867 \\
        C1.4 &  0.633 & 0.038 &  0.559 &  0.633 &  0.707 \\
        C1.5 &  0.461 & 0.040 &  0.382 &  0.461 &  0.539 \\
        C2.1 & -1.303 & 0.085 & -1.471 & -1.303 & -1.136 \\
        C2.2 & -1.521 & 0.095 & -1.706 & -1.521 & -1.335 \\
        C3.1 & -0.485 & 0.059 & -0.600 & -0.485 & -0.369 \\
        C3.2 & -0.517 & 0.060 & -0.634 & -0.517 & -0.400 \\
        \midrule
        \textbf{Hyperparameters} & & & & & \\
        \midrule
        Precision for Random Effect & 1.33 & 0.611 &       0.474  & 1.22  & 2.83 \\
        Range for GP & 56.72 &  7.684 &      43.406  & 56.11 & 73.60\\
        Std. for GP & 2.05 & 0.259 &      1.598 & 2.03  & 2.61\\
        \bottomrule
    \end{tabular}
    \caption{Posterior summaries (mean, standard deviation, quantiles) for the intercept, regression coefficients,  campaign-specific effects, the precision for the latter, and the parameters associated with the GP of the selected model using the k-fold Cross-Validation procedure. Recalling \hyperref[sec:2]{Section~\ref*{sec:2}} P. \textit{oceanica} effect is indicated as $\gamma$ in \ref{eq:intensity}. }

    \label{tab:posterior_summaries}
\end{table}

% --- BIBLIOGRAFIA ---
\bibliographystyle{plainnat}  % Stile compatibile con arXiv
\bibliography{references}     % Il file .bib deve chiamarsi references.bib

\begin{thebibliography}{38}
\providecommand{\natexlab}[1]{#1}
\providecommand{\url}[1]{\texttt{#1}}
\expandafter\ifx\csname urlstyle\endcsname\relax
  \providecommand{\doi}[1]{doi: #1}\else
  \providecommand{\doi}{doi: \begingroup \urlstyle{rm}\Url}\fi

\bibitem[Bachl et~al.(2019)Bachl, Lindgren, Borchers, and Illian]{bachl2019inlabru}
Fabian~E. Bachl, Finn Lindgren, David~L. Borchers, and Janine~B. Illian.
\newblock {inlabru}: an {R} package for {Bayesian} spatial modelling from ecological survey data.
\newblock \emph{Methods in Ecology and Evolution}, 10:\penalty0 760--766, 2019.
\newblock \doi{10.1111/2041-210X.13168}.

\bibitem[Baddeley et~al.(2005)Baddeley, Turner, M{\o}ller, and Hazelton]{baddeley2005residual}
Adrian Baddeley, Rolf Turner, Jesper M{\o}ller, and Martin Hazelton.
\newblock Residual analysis for spatial point processes (with discussion).
\newblock \emph{Journal of the Royal Statistical Society Series B: Statistical Methodology}, 67\penalty0 (5):\penalty0 617--666, 2005.

\bibitem[Berman and Turner(1992)]{berman1992approximating}
Mark Berman and T~Rolf Turner.
\newblock Approximating point process likelihoods with glim.
\newblock \emph{Journal of the Royal Statistical Society: Series C (Applied Statistics)}, 41\penalty0 (1):\penalty0 31--38, 1992.

\bibitem[Bompiani et~al.(2020)Bompiani, Petrillo, Jona~Lasinio, and Palini]{9251143}
Edoardo Bompiani, Umberto~Ferraro Petrillo, Giovanna Jona~Lasinio, and Francesco Palini.
\newblock High-performance computing with terastat.
\newblock In \emph{2020 IEEE Intl Conf on Dependable, Autonomic and Secure Computing, Intl Conf on Pervasive Intelligence and Computing, Intl Conf on Cloud and Big Data Computing, Intl Conf on Cyber Science and Technology Congress (DASC/PiCom/CBDCom/CyberSciTech)}, pages 499--506, 2020.
\newblock \doi{10.1109/DASC-PICom-CBDCom-CyberSciTech49142.2020.00092}.

\bibitem[Brix and Diggle(2001)]{brix2001spatiotemporal}
Anders Brix and Peter~J Diggle.
\newblock Spatiotemporal prediction for log-gaussian cox processes.
\newblock \emph{Journal of the Royal Statistical Society: Series B (Statistical Methodology)}, 63\penalty0 (4):\penalty0 823--841, 2001.

\bibitem[Casoli et~al.(2017)Casoli, Ventura, Cutroneo, Capello, Jona-Lasinio, Rinaldi, Criscoli, Belluscio, and Ardizzone]{casoli2017assessment}
E~Casoli, D~Ventura, L~Cutroneo, M~Capello, G~Jona-Lasinio, R~Rinaldi, A~Criscoli, A~Belluscio, and GD~Ardizzone.
\newblock Assessment of the impact of salvaging the costa concordia wreck on the deep coralligenous habitats.
\newblock \emph{Ecological Indicators}, 80:\penalty0 124--134, 2017.

\bibitem[Cronie et~al.(2024)Cronie, Moradi, and Biscio]{cronie2024cross}
Ottmar Cronie, Mehdi Moradi, and Christophe~AN Biscio.
\newblock A cross-validation-based statistical theory for point processes.
\newblock \emph{Biometrika}, 111\penalty0 (2):\penalty0 625--641, 2024.

\bibitem[Fallati et~al.(2024)Fallati, Panieri, Argentino, Varzi, B{\"u}nz, and Savini]{fallati2024combining}
Luca Fallati, Giuliana Panieri, Claudio Argentino, Andrea~Giulia Varzi, Stefan B{\"u}nz, and Alessandra Savini.
\newblock Combining rov-based acoustic data and underwater photogrammetry to characterize hakon mosby mud volcano (barents sea) cold seep systems.
\newblock In \emph{EGU General Assembly Conference Abstracts}, page 10620, 2024.

\bibitem[Gelfand and Schliep(2018)]{gelfand2018bayesian}
Alan~E Gelfand and Erin~M Schliep.
\newblock Bayesian inference and computing for spatial point patterns.
\newblock In \emph{NSF-CBMS regional conference series in probability and statistics}, volume~10, pages i--125. JSTOR, 2018.

\bibitem[Gneiting and Raftery(2007)]{gneiting2007strictly}
Tilmann Gneiting and Adrian~E Raftery.
\newblock Strictly proper scoring rules, prediction, and estimation.
\newblock \emph{Journal of the American statistical Association}, 102\penalty0 (477):\penalty0 359--378, 2007.

\bibitem[Gonz{\'a}lez-Wang{\"u}emert et~al.(2014)Gonz{\'a}lez-Wang{\"u}emert, Aydin, and Conand]{gonzalez2014assessment}
Mercedes Gonz{\'a}lez-Wang{\"u}emert, Mehmet Aydin, and Chantal Conand.
\newblock Assessment of sea cucumber populations from the aegean sea (turkey): First insights to sustainable management of new fisheries.
\newblock \emph{Ocean \& Coastal Management}, 92:\penalty0 87--94, 2014.

\bibitem[Hamel et~al.(2022)Hamel, Eeckhaut, Conand, Sun, Caulier, and Mercier]{hamel2022global}
Jean-Fran{\c{c}}ois Hamel, Igor Eeckhaut, Chantal Conand, Jiamin Sun, Guillaume Caulier, and Annie Mercier.
\newblock Global knowledge on the commercial sea cucumber holothuria scabra.
\newblock \emph{Advances in marine biology}, 91:\penalty0 1--286, 2022.

\bibitem[Jordan et~al.(2019)Jordan, Kr{\"u}ger, and Lerch]{jordan2019evaluating}
Alexander Jordan, Fabian Kr{\"u}ger, and Sebastian Lerch.
\newblock Evaluating probabilistic forecasts with scoringrules.
\newblock \emph{Journal of Statistical Software}, 90:\penalty0 1--37, 2019.

\bibitem[Kr{\"u}ger et~al.(2021)Kr{\"u}ger, Lerch, Thorarinsdottir, and Gneiting]{kruger2021predictive}
Fabian Kr{\"u}ger, Sebastian Lerch, Thordis Thorarinsdottir, and Tilmann Gneiting.
\newblock Predictive inference based on markov chain monte carlo output.
\newblock \emph{International Statistical Review}, 89\penalty0 (2):\penalty0 274--301, 2021.

\bibitem[Leininger and Gelfand(2017)]{leininger2017bayesian}
Thomas~J Leininger and Alan~E Gelfand.
\newblock Bayesian inference and model assessment for spatial point patterns using posterior predictive samples.
\newblock \emph{Bayesian Analysis}, 12\penalty0 (1):\penalty0 1--30, 2017.

\bibitem[Lindgren et~al.(2011)Lindgren, Rue, and Lindström]{lindgren2011gmrf}
Finn Lindgren, Håvard Rue, and Johan Lindström.
\newblock An explicit link between gaussian fields and gaussian markov random fields: the stochastic partial differential equation approach.
\newblock \emph{Journal of the Royal Statistical Society: Series B (Statistical Methodology)}, 73\penalty0 (4):\penalty0 423--498, 2011.
\newblock \doi{https://doi.org/10.1111/j.1467-9868.2011.00777.x}.
\newblock URL \url{https://rss.onlinelibrary.wiley.com/doi/abs/10.1111/j.1467-9868.2011.00777.x}.

\bibitem[Lopez and Levinton(1987)]{lopez1987ecology}
Glenn~R Lopez and Jeffrey~S Levinton.
\newblock Ecology of deposit-feeding animals in marine sediments.
\newblock \emph{The quarterly review of biology}, 62\penalty0 (3):\penalty0 235--260, 1987.

\bibitem[Mancini et~al.(2019)Mancini, Casoli, Ventura, Jona-Lasinio, Criscoli, Belluscio, and Ardizzone]{mancini2019impact}
G~Mancini, E~Casoli, D~Ventura, G~Jona-Lasinio, A~Criscoli, A~Belluscio, and GD~Ardizzone.
\newblock Impact of the costa concordia shipwreck on a posidonia oceanica meadow: A multi-scale assessment from a population to a landscape level.
\newblock \emph{Marine Pollution Bulletin}, 148:\penalty0 168--181, 2019.

\bibitem[Mancini et~al.(2022)Mancini, Ventura, Casoli, Belluscio, and Ardizzone]{mancini2022transplantation}
G~Mancini, D~Ventura, E~Casoli, A~Belluscio, and GD~Ardizzone.
\newblock Transplantation on a posidonia oceanica meadow to facilitate its recovery after the concordia shipwrecking.
\newblock \emph{Marine Pollution Bulletin}, 179:\penalty0 113683, 2022.

\bibitem[Martino et~al.(2021)Martino, Pace, Moro, Casoli, Ventura, Frachea, Silvestri, Arcangeli, Giacomini, Ardizzone, et~al.]{martino2021integration}
Sara Martino, Daniela~Silvia Pace, Stefano Moro, Edoardo Casoli, Daniele Ventura, Alessandro Frachea, Margherita Silvestri, Antonella Arcangeli, Giancarlo Giacomini, Giandomenico Ardizzone, et~al.
\newblock Integration of presence-only data from several sources: a case study on dolphins' spatial distribution.
\newblock \emph{Ecography}, 44\penalty0 (10):\penalty0 1533--1543, 2021.

\bibitem[Mastrantonio et~al.(2024)Mastrantonio, Ventura, Casoli, Rakaj, Jona~Lasinio, Poggio, Vitiello, and Calculli]{mastrantonio2024species}
Gianluca Mastrantonio, Daniele Ventura, Edoardo Casoli, Arnold Rakaj, Giovanna Jona~Lasinio, Daniele Poggio, Cecilia Vitiello, and Crescenza Calculli.
\newblock Species distribution models with masking: The case of holothurians in a posidonia rich area.
\newblock In \emph{Scientific Meeting of the Italian Statistical Society}, pages 531--536. Springer, 2024.

\bibitem[Matheson and Winkler(1976)]{matheson1976scoring}
James~E Matheson and Robert~L Winkler.
\newblock Scoring rules for continuous probability distributions.
\newblock \emph{Management science}, 22\penalty0 (10):\penalty0 1087--1096, 1976.

\bibitem[McElreath(2018)]{mcelreath2018statistical}
Richard McElreath.
\newblock \emph{Statistical rethinking: A Bayesian course with examples in R and Stan}.
\newblock Chapman and Hall/CRC, 2018.

\bibitem[M{\o}ller et~al.(1998)M{\o}ller, Syversveen, and Waagepetersen]{moller1998log}
Jesper M{\o}ller, Anne~Randi Syversveen, and Rasmus~Plenge Waagepetersen.
\newblock Log gaussian cox processes.
\newblock \emph{Scandinavian journal of statistics}, 25\penalty0 (3):\penalty0 451--482, 1998.

\bibitem[Pasquini et~al.(2022)Pasquini, Porcu, Marongiu, Follesa, Giglioli, and Addis]{pasquini2022new}
Viviana Pasquini, Cristina Porcu, Martina~Francesca Marongiu, Maria~Cristina Follesa, Ambra~Angelica Giglioli, and Pierantonio Addis.
\newblock New insights upon the reproductive biology of the sea cucumber holothuria tubulosa (echinodermata, holothuroidea) in the mediterranean: Implications for management and domestication.
\newblock \emph{Frontiers in Marine Science}, 9:\penalty0 1029147, 2022.

\bibitem[Purcell et~al.(2016)Purcell, Conand, Uthicke, and Byrne]{purcell2016ecological}
Steven~W Purcell, Chantal Conand, Sven Uthicke, and Maria Byrne.
\newblock Ecological roles of exploited sea cucumbers.
\newblock In \emph{Oceanography and marine biology}, pages 375--394. CRC press, 2016.

\bibitem[Rakaj and Fianchini(2024)]{rakaj2024mediterranean}
Arnold Rakaj and Alessandra Fianchini.
\newblock Mediterranean sea cucumbers—biology, ecology, and exploitation.
\newblock In \emph{The world of sea cucumbers}, pages 753--773. Elsevier, 2024.

\bibitem[Rue et~al.(2009)Rue, Martino, and Chopin]{rue2009inla}
Håvard Rue, Sara Martino, and Nicolas Chopin.
\newblock Approximate bayesian inference for latent gaussian models by using integrated nested laplace approximations.
\newblock \emph{Journal of the Royal Statistical Society: Series B (Statistical Methodology)}, 71\penalty0 (2):\penalty0 319--392, 2009.
\newblock \doi{https://doi.org/10.1111/j.1467-9868.2008.00700.x}.
\newblock URL \url{https://rss.onlinelibrary.wiley.com/doi/abs/10.1111/j.1467-9868.2008.00700.x}.

\bibitem[Schneider et~al.(2013)Schneider, Silverman, Kravitz, Rivlin, Schneider-Mor, Barbosa, Byrne, and Caldeira]{schneider2013inorganic}
Kenneth Schneider, Jacob Silverman, Ben Kravitz, Tanya Rivlin, Aya Schneider-Mor, Sergio Barbosa, Maria Byrne, and Ken Caldeira.
\newblock Inorganic carbon turnover caused by digestion of carbonate sands and metabolic activity of holothurians.
\newblock \emph{Estuarine, Coastal and Shelf Science}, 133:\penalty0 217--223, 2013.

\bibitem[Serra et~al.(2014)Serra, Saez, Mateu, Varga, Juan, D{\'\i}az-{\'A}valos, and Rue]{serra2014spatio}
Laura Serra, Marc Saez, Jorge Mateu, Diego Varga, Pablo Juan, Carlos D{\'\i}az-{\'A}valos, and H{\aa}vard Rue.
\newblock Spatio-temporal log-gaussian cox processes for modelling wildfire occurrence: the case of catalonia, 1994--2008.
\newblock \emph{Environmental and ecological statistics}, 21:\penalty0 531--563, 2014.

\bibitem[Sicacha-Parada et~al.(2021)Sicacha-Parada, Steinsland, Cretois, and Borgelt]{sicacha2021accounting}
Jorge Sicacha-Parada, Ingelin Steinsland, Benjamin Cretois, and Jan Borgelt.
\newblock Accounting for spatial varying sampling effort due to accessibility in citizen science data: A case study of moose in norway.
\newblock \emph{Spatial Statistics}, 42:\penalty0 100446, 2021.

\bibitem[Simpson et~al.(2016)Simpson, Illian, Lindgren, S{\o}rbye, and Rue]{simpson2016going}
Daniel Simpson, Janine~Baerbel Illian, Finn Lindgren, Sigrunn~H S{\o}rbye, and Havard Rue.
\newblock Going off grid: Computationally efficient inference for log-gaussian cox processes.
\newblock \emph{Biometrika}, 103\penalty0 (1):\penalty0 49--70, 2016.

\bibitem[Simpson et~al.(2017)Simpson, Rue, Riebler, Martins, and S{\o}rbye]{simpson2017penalising}
Daniel Simpson, H{\aa}vard Rue, Andrea Riebler, Thiago~G Martins, and Sigrunn~H S{\o}rbye.
\newblock Penalising model component complexity: a principled, practical approach to constructing priors.
\newblock \emph{Statistical Science}, 32\penalty0 (1):\penalty0 1--28, 2017.

\bibitem[Toniolo et~al.(2018)Toniolo, Di~Sotto, Di~Giacomo, Ventura, Casoli, Belluscio, Nicoletti, and Ardizzone]{toniolo2018seagrass}
Chiara Toniolo, Antonella Di~Sotto, Silvia Di~Giacomo, Daniele Ventura, Edoardo Casoli, Andrea Belluscio, Marcello Nicoletti, and Giandomenico Ardizzone.
\newblock Seagrass posidonia oceanica (l.) delile as a marine biomarker: a metabolomic and toxicological analysis.
\newblock \emph{Ecosphere}, 9\penalty0 (3):\penalty0 e02054, 2018.

\bibitem[Ventura et~al.(2022)Ventura, Castoro, Mancini, Casoli, Pace, Belluscio, and Ardizzone]{ventura2022high}
Daniele Ventura, Luca Castoro, Gianluca Mancini, Edoardo Casoli, Daniela~Silvia Pace, Andrea Belluscio, and Giandomenico Ardizzone.
\newblock High spatial resolution underwater data for mapping seagrass transplantation: A powerful tool for visualization and analysis.
\newblock \emph{Data in Brief}, 40:\penalty0 107735, 2022.

\bibitem[Ventura et~al.(2025)Ventura, Rakaj, Lasinio, Sangiovanni, Poggio, Mastrantonio, Pollice, Grasso, Casoli, Mancini, et~al.]{ventura2025detecting}
Daniele Ventura, Arnold Rakaj, Giovanna~Jona Lasinio, Gian~Mario Sangiovanni, Daniele Poggio, Gianluca Mastrantonio, Alessio Pollice, Gaia Grasso, Edoardo Casoli, Gianluca Mancini, et~al.
\newblock Detecting habitat preferences and monitoring population dynamics of sea cucumbers in coastal ecosystems through underwater photogrammetry.
\newblock \emph{Journal of Environmental Management}, 377:\penalty0 124589, 2025.

\bibitem[Warton and Shepherd(2010)]{warton2010poisson}
David~I Warton and Leah~C Shepherd.
\newblock Poisson point process models solve the" pseudo-absence problem" for presence-only data in ecology.
\newblock \emph{The Annals of Applied Statistics}, pages 1383--1402, 2010.

\bibitem[Yuan et~al.(2017)Yuan, Bachl, Lindgren, Borchers, Illian, Buckland, Rue, and Gerrodette]{yuan2017point}
Yuan Yuan, Fabian~E Bachl, Finn Lindgren, David~L Borchers, Janine~B Illian, Stephen~T Buckland, H{\aa}vard Rue, and Tim Gerrodette.
\newblock Point process models for spatio-temporal distance sampling data from a large-scale survey of blue whales.
\newblock \emph{Annals of Applied Statistics}, 11\penalty0 (4):\penalty0 2270--2297, 2017.

\end{thebibliography}

\end{document}